# AN AIDING TOOL FOR BUILDING DESIGN GENERATION, THERMAL ASSESSMENT AND OPTIMIZATION – ENERGYPLUS INTERACTION OVERVIEW

**Marco S. Fernandes[1] \*, Eugénio Rodrigues[1], Adélio Rodrigues Gaspar[1], Álvaro Gomes[2]**

1: ADAI, LAETA, Department of Mechanical Engineering, University of Coimbra
Rua Luís Reis Santos, Pólo II, 3030-788 Coimbra, Portugal, marco.fernandes@adai.pt,
eugenio.rodrigues@gmail.com, adelio.gaspar@dem.uc.pt

2: INESC Coimbra, Department of Electrical and Computer Engineering, University of Coimbra
Rua Sílvio Lima, Pólo II, 3030-290 Coimbra, Portugal, agomes@deec.uc.pt

**Abstract**: *A building design aiding tool for space allocation and thermal performance optimization is being developed to help practitioners during the building space planning phase, predicting how it will behave regarding energy consumption and thermal comfort. The tool evaluates, ranks, and optimizes generated floor plans according to thermal performance criteria, using the dynamic simulation program EnergyPlus. The tool is currently able to use a wide variety of EnergyPlus objects, allowing for various template and detailed HVAC, DHW, and thermal and electrical energy production systems and components, as well as numerous internal gains types, construction elements and energy saving controls, to be accounted for and simulated in the generated buildings. This paper presents the tool overall concept as well as the main features regarding dynamic simulation. Some performance results are presented for distinct systems to illustrate the use and potential of the tool.*

## 1. INTRODUCTION

The efficient use of energy in the buildings sector is a significant issue for a more sustainable future. Currently, buildings account for a third of the global energy consumption, with space heating and cooling representing roughly half of that value [1]. However, this end-use consumption can be reduced if buildings have more adequate designs. There are already several simulation software tools capable of assessing the building's design performance in numerous fields, such as energy consumption, visual comfort, indoor air quality and thermal comfort, etc. In an ideal scenario, those should be included in the building design process, since energy efficient solutions are easier to incorporate at the early design phases. However, as it requires a long time to build accurate simulation models and to understand which options are more adequate in each building design stage, those tools are not widely used, which makes difficult to predict and incorporate energy efficiency strategies [2]. In this context, this paper presents an aiding design tool for the space allocation and thermal performance optimization, which evaluates, ranks, and optimizes the generated floor plans according to thermal comfort or energy consumption criteria, using the coupled dynamic simulation software EnergyPlus. This tool is being extended to help practitioners during the deep building renovation scenarios. It allows to evaluate and compare the performance of a large number of alternative solutions [3] or even to improve those solutions with optimization techniques [4], and provides useful energy performance data or thermal improvements based on space organization, thus assisting the practitioners in their decision-making process. Currently, a wide variety of EnergyPlus objects are implemented in the tool, allowing for several equipment, systems and usage objects to be simulated: HVAC, DHW, and thermal and electrical energy production systems and components, internal gains objects, materials and construction elements, usage schedules, among others.

This paper presents the tool overall concept, focusing on the main features regarding dynamic simulation and the EnergyPlus functionalities already implemented. In the end, some performance results are presented for distinct systems to exemplify the use and potential of the tool.

## 2. TOOL DESCRIPTION

The tool workflow starts by setting the user's preferences and requirements (Layout Specifications Program module – LSP). After, the Evolutionary Program for the Space Allocation Program (EPSAP) algorithm [5–7] will generate the desired alternative building geometries according to those preferences and requirements. The best solutions are then evaluated for the first time according to the performance criteria (Building Performance Simulation module – BPS) set by the user. Depending on the criteria, a different performance engine is chosen: the thermal comfort and the energy consumption criteria will invoke EnergyPlus and the visual comfort criteria will use Radiance simulation engine. Before the solutions are presented to the user in the Graphical User Interface (GUI) module, the building geometry can be optimized using the Floor Plan Performance Optimization Program (FPOP) algorithm [4]. See Figure 1 for the complete workflow structure.

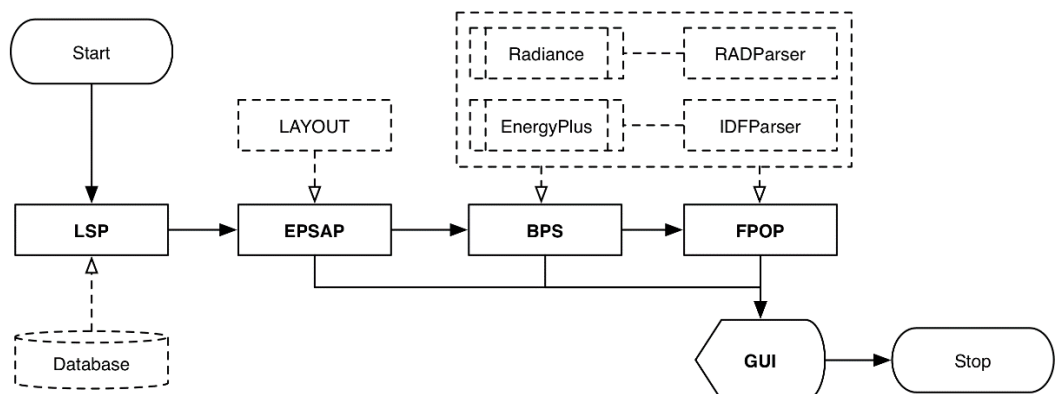

Figure 1. The tool workflow.

## 2.1. Design generation

The tool generates alternative building designs according to the user's preferences and requirements. The generative design method used to generate the space arrangements is the EPSAP algorithm, presented in refs. [5–7], which is a two-stage hybrid evolutionary strategy technique that takes into account a set of objectives identified by the user and technical or user defined constraints. A set of performance indicators are aggregated in an objective function assessing the quality of solutions computed using a hybrid evolution strategy. The architectural elements are subject to geometric transformations, such as translation, rotation, stretching, mirroring, etc., aiming to improve the performance indicators. These transformations are applied to single objects (openings and spaces), clusters of objects, storeys, or to the whole building. Figure 2 presents examples of alternative design solutions generated for single, two, and three-storey family houses.

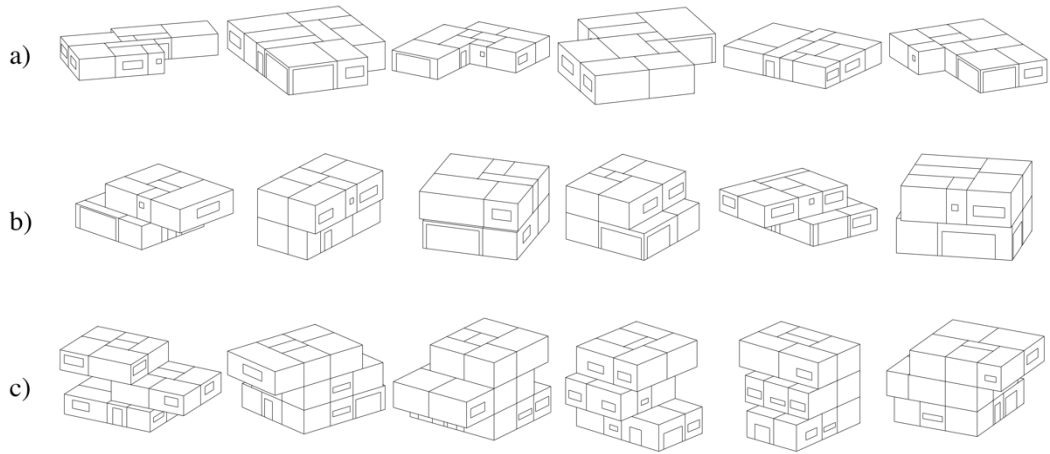

Figure 2. Examples of generated building for single-storey (a), two-storey (b), and three-storey family house (c).

## 2.2. Performance simulation

When the floor plan generation is complete, the energy consumption or the thermal comfort of the generated solutions can be evaluated using the coupled dynamic simulation engine – EnergyPlus [8] – according to the selected performance objective criteria [4,9]. This evaluates the influence of a set of variables, such as the building constructive elements (walls, ceilings, roofs, pavements, windows, and doors), location (with respective weather data), internal gains (people, lighting, and electric, gas, steam and water equipment), ventilation and infiltration, HVAC, DHW and electric generation systems, usage schedules, among others, which are specified according to default data from a database and existing templates, or specifically set by the user. After the first simulation and definition of the performance criterium (BPS module), the building's performance can be improved according to the user design strategy [4], using a sequential variable optimization procedure (the FPOP algorithm [4]) that changes the geometry and/or adds shading elements to the building design, thus minimizing a weighted sum of cooling and heating degree-hours of thermal discomfort in all spaces.

The simulation results are then presented in output files by system, space and meter type, and in a graphical user interface (GUI) that displays the building floor plans and the overall and space performance reports – Figure 3. The GUI displays the best of the generated layouts (though the user can visualize other layouts and storeys, if available) and a set of performance reports, as well as other results, such as the building areas and the building and systems costs. The performance reports illustrate graphically the dynamic simulation results for the whole building and for its individual spaces, through a set of variables that are accessible for user selection depending on the simulated systems: *e.g.*, indoor and outdoor air temperatures, cooling and heating energy, infiltration rate, water consumption, primary energy consumption, thermal discomfort, electric equipment energy consumption. Moreover, the user can also select the report period for which each graphic will be generated: all year, trimester, coldest day, or hottest day.

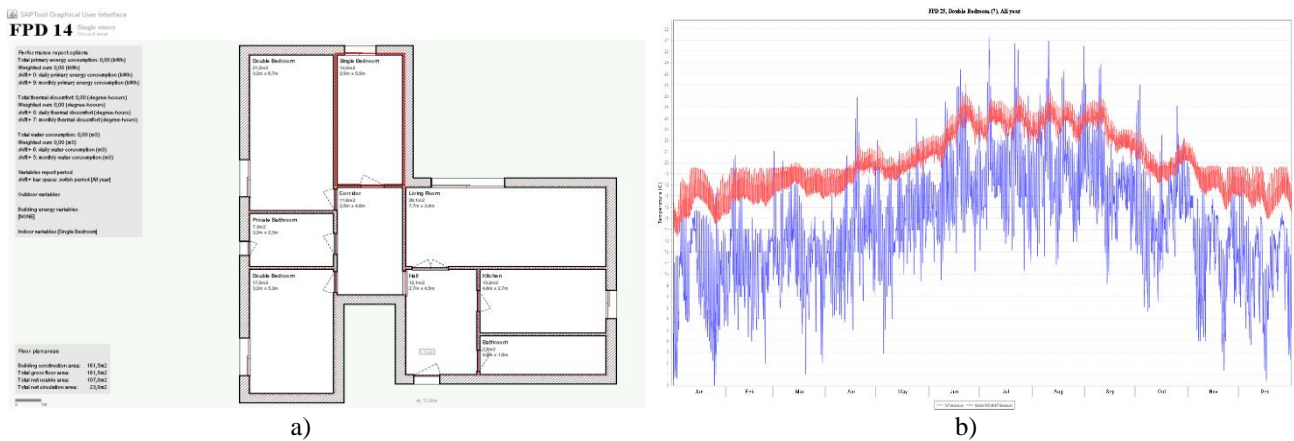

a)                                                      b)

Figure 3. a) Graphical User Interface; b) Annual air temperature graphic generated for one of the building zones.

## 3.  MAIN FEATURES IMPLEMENTED FOR DYNAMIC SIMULATION

Currently, the following EnergyPlus main functionalities are implemented in the tool for dynamic simulation, according to the latest specifications (version 8.8.0) [10]:

- User defined materials and constructions for the building surfaces and subsurfaces;
- Automatic definition of thermal zones for the building spaces generated by the tool;
- Allocation of user defined internal gains (people, lighting, and equipment) to the thermal zones;
- Optional window shading and daylighting controls;
- Simple zone ventilation and infiltration and detailed airflow network options are available;
- Forced and natural ventilation options;
- Cold/hot water systems: terminal zone equipment, equipment connections, storage tanks;
- Various HVAC system templates: ideal loads air system, baseboard heat, zone unitary system and terminal units, hot water plant loop;
- Various zone HVAC detailed systems: ideal loads air system, low temperature radiant floor, baseboard water convective, baseboard electric, air terminal VAV reheat, air terminal uncontrolled;
- Several HVAC components: pumps, pipes, fans, humidifiers, heat exchangers, heating and cooling coils, solar collector flat plate, district heating and cooling, boilers, water heater tank;
- User defined plant loops (hot/cold water, steam), using the HVAC components, to feed zone HVAC water/steam systems and/or DHW equipment;
- User defined air loops, using the HVAC components, to feed zone HVAC air systems;
- User defined electric load center, comprising inverters, batteries, transformers, and photovoltaic and/or wind turbine generators.

In addition, the tool presents several useful features regarding the EnergyPlus interaction:

- Various buildings and systems templates are available;
- Object values (*e.g.*, properties, setpoints) are assigned in a specific database;
- Detailed systems are easy to define;
- Simple connection between different component types;
- All equipment and system nodes are automatically defined;
- Most object names are automatically defined, depending on the equipment, systems and zones for which they are assigned;
- Possibility to define equipment and construction costs, which are then automatically sorted by type in the GUI.

## 4. RESULTS

The following figures depict some results of the BPS module simulation process displayed in the GUI, in which the user has access to both the generated plants and the output reports of the simulated systems: Figure 4 – DHW plant loop system comprising a water tank with a solar collectors loop in its source side and zone water equipment in its use side; Figure 5 – electric load center comprising photovoltaic and wind turbine generators, an inverter, and a storage battery; and Figure 6 – air loop HVAC comprising an unitary system with heating and cooling coils for zone climatization.

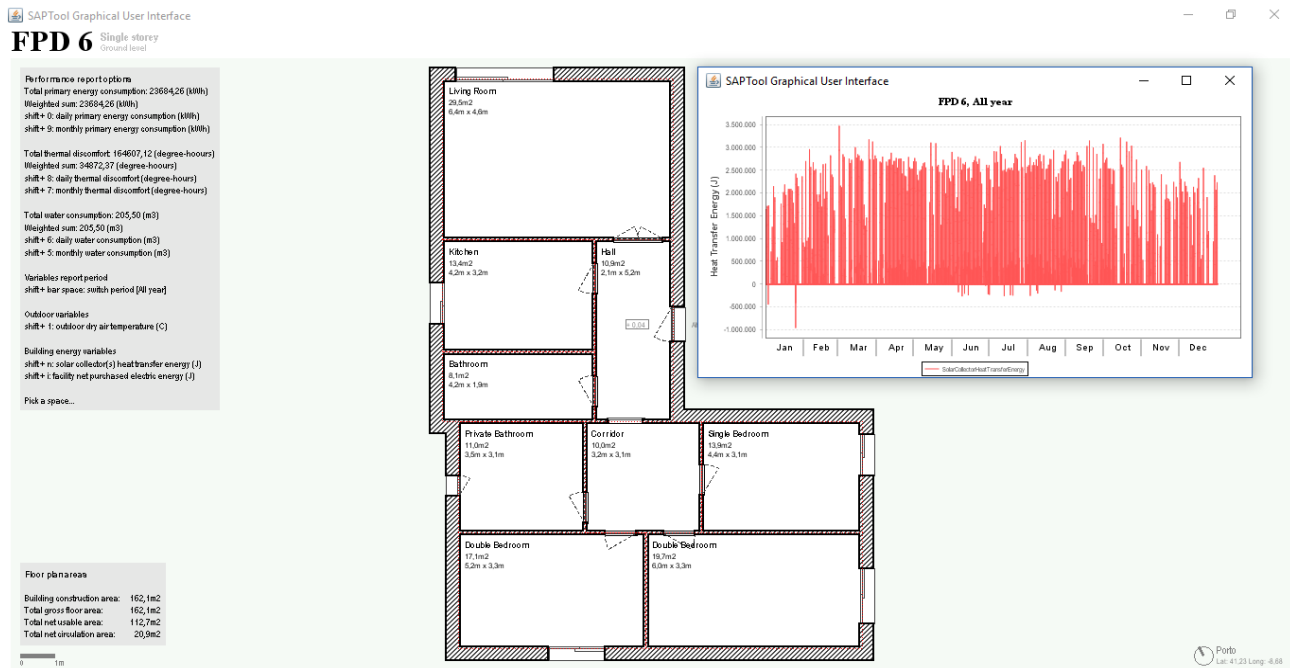

Figure 4. Building design example, and graphic depicting solar collector heat transfer energy for whole year (DHW plant loop).

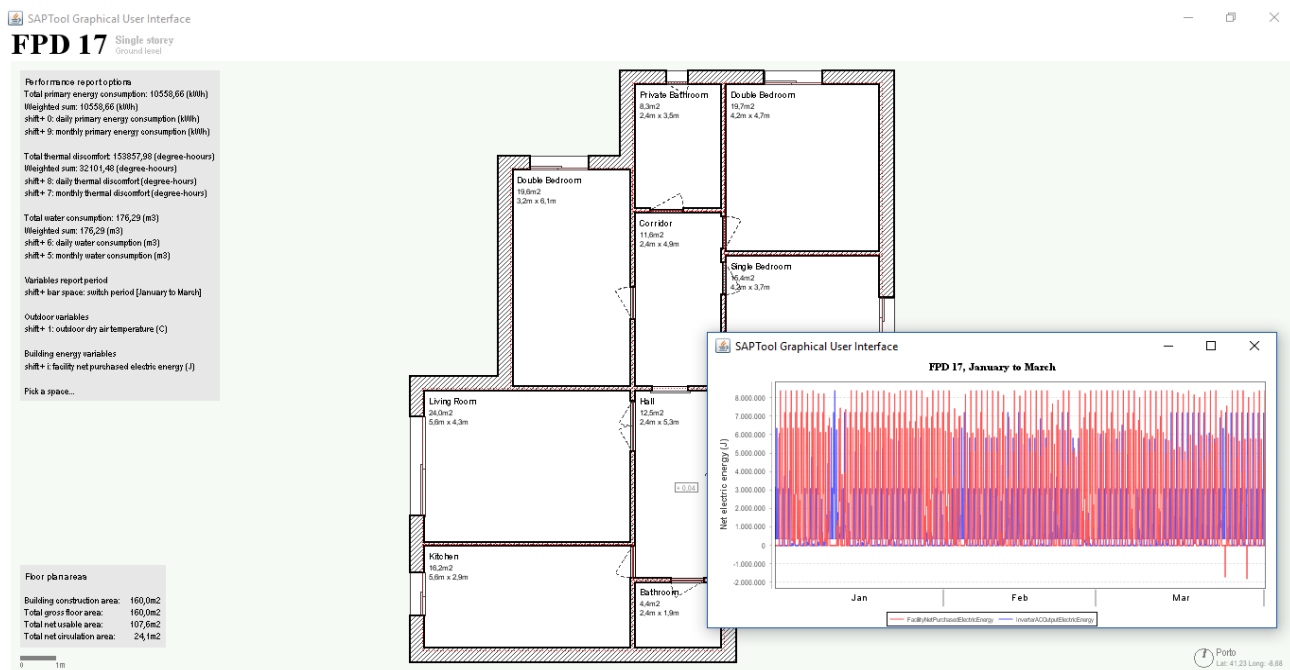

Figure 5. Building design example, and graphic depicting the net purchased and the produced (electric load center generators) electric energy in the first trimester.

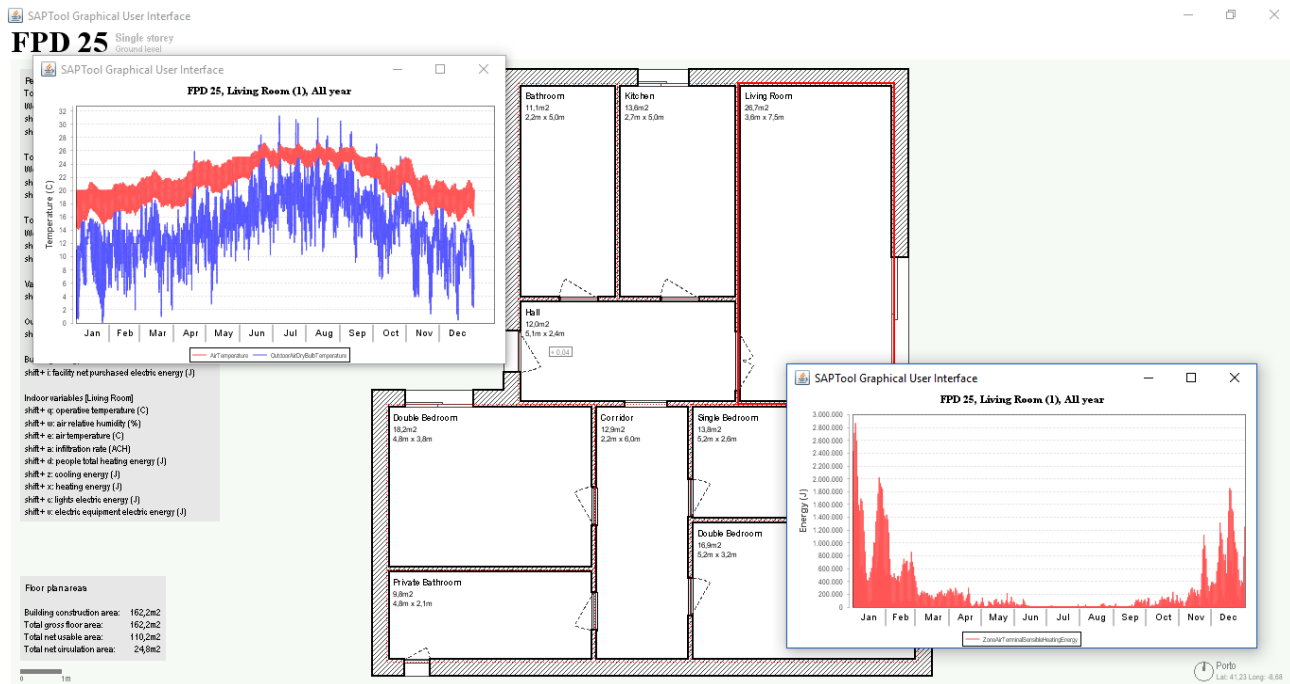

Figure 6. Building design example, and graphics depicting zone indoor and outdoor dry-bulb temperatures and heating energy for whole year (air loop HVAC with unitary system).

The available results facilitate the practitioners' decision-making process, as they display the best generated layouts according to the user's initial preferences and requirements, as well as the performance outputs for both the building and its systems; thus, allowing to quickly select the best performance solution(s) for a given typology, or even to compare the building's performance for different layouts and/or systems.

## 5. CONCLUSIONS

Building performance simulation software can contribute significantly to the improvement of buildings' energy efficiency and thermal comfort by supporting the decision-making process. The current tools are, however, difficult to implement in the building design process, as they usually increase the effort of an already long architectural design methodology. In order to overcome this deficit, the tool presented in this paper automatically combines the generation, evaluation and optimization of alternative design solutions in the space planning phase. It offers a wide variety of EnergyPlus objects, thus allowing to simulate several internal gain types, water systems, template and detailed HVAC systems, electrical generation systems, and individual components. These are semi-automatically defined, sparing the user from complex tasks, such as naming, assignment and linkage of systems and equipment nodes. In addition, a simple graphical user interface displays the main building and space performance reports in a user-friendly environment. Therefore, this aiding tool can offer building practitioners a valuable interactive decision support appliance. Its current results show already great added value, by allowing to interact space generation with energy dynamic simulation, thus facilitating the decision-making process in different building design stages. Additionally, the tool is also a valuable scientific instrument to study specific building performance aspects by producing parametric analysis of specific decision variables [11] or create large datasets of different kinds of buildings to be statistically analysed [3,12]. The tool is currently being extended to implement more EnergyPlus objects, in order to improve its simulation capabilities and to offer a more realistic approach to building usage and systems dynamics. This will allow to size and optimize the system according to the building type and adjust the building geometry to match the energy systems, using the FPOP module.

## ACKNOWLEDGEMENTS

This research work is supported by the Portuguese Foundation for Science and Technology (FCT) and European Regional Development Fund (FEDER) through COMPETE with references PTDC/EMS-ENE/3238/2014, POCI-01-0145-FEDER-016760, and LISBOA-01-0145-FEDER-016760. Eugénio Rodrigues acknowledges the support provided by the FCT, under PostDoc grant SFRH/BPD/99668/2014.